\documentclass[10pt]{article}
\usepackage{euscript,amssymb,amsmath}
\usepackage[dvips]{graphicx}
\catcode`\@=11

\newcommand{\be}[1]{\begin{equation}\label{#1}}
\newcommand{\ee}{\end{equation}}
\newcommand{\ba}[1]{\begin{eqnarray}\label{#1}}
\newcommand{\ea}{\end{eqnarray}}
\newcommand{\rf}[1]{(\ref{#1})}
\newcommand{\nn}{\nonumber}

\title{How to play a disc brake
\vspace{5mm}
\\
\large{O.N. Kirillov}
\\
\vspace{3mm} {\small Institute of Mechanics, Moscow State Lomonosov
University, Michurinskii pr. 1, 119192 Moscow, Russia
E-mail:~kirillov@imec.msu.ru} }
\date{}
\begin{document}
\maketitle
\vspace{-10mm}
\begin{abstract}
We consider a gyroscopic system with two degrees of freedom under
the action of small dissipative and non-conservative positional forces, which has its origin
in the models of rotating bodies of revolution being in frictional contact.
The spectrum of the unperturbed gyroscopic system forms a "spectral
mesh" in the plane "frequency -- gyroscopic parameter" with double
semi-simple purely imaginary eigenvalues at zero value of the gyroscopic parameter.
It is shown that dissipative forces lead to the splitting of the
semi-simple eigenvalue with the creation of the so-called "bubble of
instability" -- a ring in the three-dimensional space of the gyroscopic
parameter and real and imaginary parts of eigenvalues, which corresponds to complex
eigenvalues. In case of full dissipation with a positive-definite damping matrix
the eigenvalues of the ring have negative real parts making the bubble a latent source
of instability because it can "emerge" to the region of eigenvalues with positive real
parts due to action of both indefinite damping and non-conservative positional forces.
In the paper, the instability mechanism is analytically described with the use
of the perturbation theory of multiple eigenvalues. Explicit conditions are established
for the origination of the bubble of instability and its transition from the
latent to active phase, clarifying the key role of indefinite damping
and non-conservative positional forces in the development and localization of the subcritical
flutter instability. As an example stability of a rotating circular string constrained
by a stationary load system is studied in detail. The theory developed seems to give
a first clear explanation of
the mechanism of self-excited vibrations in the rotating structures in frictional contact,
that is responsible for such well-known phenomena of acoustics of friction as
the squealing disc brake and the singing wine glass.
\end{abstract}

\begin{flushleft}
\end{flushleft}

\section{Introduction}

An axially symmetric shell, like a wine glass can easily produce
sound when a wet finger is rubbed around its rim or wall, as it was
observed already in 1638 by Galileo Galilei in his Dialogues
Concerning Two New Sciences \cite{GG1638, Fr83, S04}. This principle
is used in playing the glass harmonica invented by Benjamin Franklin
in 1757, which he called "armonica", where, in order to produce
sound, one should touch by a moist finger an edge of a glass bowl
rotating around its axis of symmetry \cite{Ro94,Ak02,S04,GA}. This
remarkable phenomenon has been studied experimentally; see e.g.
\cite{JRFDV06}. However, an adequate analytical theory for its
description seems to be still missing. Another closely related
example of acoustics of friction is the squealing disc brake
\cite{Ak02,KOP03}. This mechanical system produces sound due to
transverse vibrations of a rotating annular plate caused by its
interaction with the brake pads. Despite intensive experimental and
theoretical study, the problem of predicting and controlling the
squeal remains an important issue
\cite{IS73,PS90,Mo98,Ak02,PP02,KOP03,S04,BW06,Ha07,SHKH07}.
Significant but still poorly understood phenomena are squealing and
barring of calender rolls in paper mills causing an intensive noise
and reducing the quality of the paper \cite{H05}.

The presence of multiple eigenvalues in the spectra of free vibrations of axially symmetric
shells and plates is well-known. Already Rayleigh, studying the acoustics of bells, recognized that,
if the symmetry of a bell were complete, the nodal meridians of a transverse vibration mode would
have no fixed position but would travel freely around the bell, as do those in a wine glass driven
by the moistened finger \cite{Ak02}. This is a reflection of the fact that spectra of free vibrations of a bell,
a wine glass, an annular plate and other bodies of revolution contain double purely imaginary semi-simple
eigenvalues with two linearly independent eigenvectors.

Rotation causes the double eigenvalues of an axially symmetric structure to split \cite{Br1890}.
The newborn pair of simple eigenvalues corresponds to the forward and backward travelling waves,
which propagate along the circumferential direction \cite{Br1890,So22,IS73,Sh84,YH95,Mo98,TH99}.
Viewed from the rotational frame, the frequency of the forward travelling wave appears to
decrease and that of the backward travelling wave appears to increase, as the spin increases.
Due to this fact, double eigenvalues originate again at non-zero angular velocities,
forming the nodes of the spectral mesh in the plane 'eigenfrequency' versus 'angular velocity'.
The spectral meshes are characteristic for example for the rotating circular strings,
rings, discs, and cylindrical and hemispherical shells.
The phenomenon is apparent also in hydrodynamics, in the problem of stability of a vortex tube \cite{Fu03}
and in magnetohydrodynamics (MHD) in the problem of instability of the spherically symmetric MHD $\alpha^2$-dynamo
\cite{GK06}.

It is known that striking the wine glass excites a number of modes, but rubbing the rim with
a finger or bowing it radially with a violin bow generally excites a single mode \cite{Ro94}.
The same is true for the squealing disc brake \cite{KOP03,MGB06,GAM06,GS06,Ha07,SHKH07}. For this reason,
we formulate the main problem of acoustics of friction of rotating elastic bodies of revolution as the description
of the mechanism of activating a particular mode of the continuum by its contact with an external body.

In case of the disc brake, the frictional contact of the brake pads with the rotating disc introduces
dissipative and non-conservative positional forces into the system \cite{Ha07,SHKH07}.
Since the nodes of the spectral mesh correspond to the double eigenvalues, they are most sensitive to
perturbations, especially to those breaking the symmetries of the system. Consequently, the instability will
most likely occur at the angular velocities close to that of the nodes of the spectral mesh and the unstable
modes of the perturbed system will have the frequencies close to that of the double eigenvalues
at the nodes. This picture qualitatively agrees with the existing experimental data
\cite{Ro94,Ak02,KOP03,MGB06,GAM06,GS06}.

In the following with the use of the perturbation theory of multiple eigenvalues \cite{KS04,KMS05,KS05,GK06} we will show
that independently on the definiteness of the damping matrix, there exist combinations of
dissipative and non-conservative positional forces causing the flutter instability in the vicinity
of the nodes of the spectral mesh for the angular velocities from the subcritical range.
We will show that zero and negative eigenvalues in the spectrum of the damping matrix
encourage the development of the localized subcritical flutter instability while zero eigenvalues in the matrix
of non-conservative positional forces suppress it. Explicit expressions describing the
movements of eigenvalues due to change of the system parameters will be obtained. Conditions
will be derived for the eigenvalues to move to the right part of the complex plane.
Approximations of the domain of asymptotic stability will be found and singularities of the stability boundary
responsible for the development of instability will be described and classified. The methodology
developed for the study of the two-dimensional system will finally be employed to the detailed investigation
of the stability of a rotating circular string constrained by a stationary load system.

    \begin{figure}
    \begin{center}
    \includegraphics[angle=0, width=0.4\textwidth]{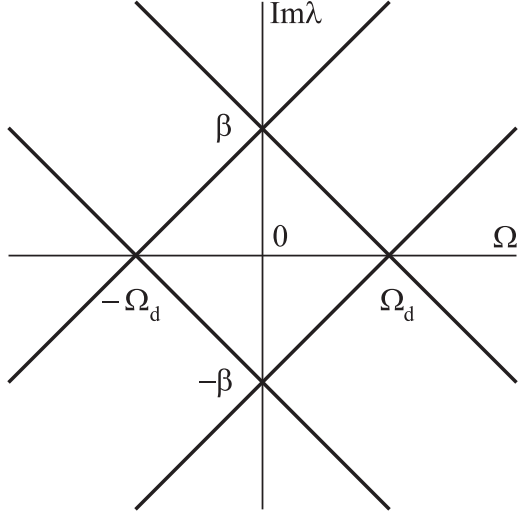}
    \end{center}
    \caption{The spectral mesh of system \rf{i1} when $\delta=\nu=\kappa=0$. }
    \label{fig1}
    \end{figure}

\section{The spectral mesh of a two-dimensional gyroscopic system}

Consider an autonomous non-conservative system described by a linear
differential equation of second order
\be{i1}
\ddot {\bf x} +
(2\Omega{\bf G}+\delta{\bf D})\dot {\bf x} + ((\beta^2-\Omega^2){\bf I}+\kappa{\bf K}+\nu{\bf
N}){\bf x} =0,
\ee
where a dot over a symbol denotes time differentiation, ${\bf
x}\in \mathbb{R}^2$, and $\bf I$ is the identity matrix.
The real matrices ${\bf D}={\bf D}^T$, ${\bf
G}=-{\bf G}^T$, ${\bf K}={\bf K}^T$, and ${\bf N}=-{\bf N}^T$ are related to dissipative
(damping), gyroscopic, potential, and non-conservative positional (circulatory)
forces with magnitudes controlled by scaling factors $\delta$,
$\Omega$, $\kappa$, and $\nu$ respectively; $\beta>0$ is the
frequency of free vibrations of the potential system corresponding to $\delta=\Omega=\kappa=\nu=0$.
The parameters and variables of the system are assumed to be non-dimensional quantities.
Without loss in generality we assume $\det{\bf G}=\det{\bf N}=1$.
Equation \rf{i1} appears as a two-mode approximation of the models
of rotating bodies of revolution in frictional contact
after their linearization and discretization \cite{Mo98,KOP03,SHKH07}.

Separating time by setting ${\bf x}(t)={\bf u}\exp(\lambda t)$ we arrive at the eigenvalue problem
\be{i2}
{\bf L}{\bf u}=0,\quad
{\bf L}={\bf I}\lambda^2+(2\Omega{\bf G}+\delta{\bf D})\lambda + (\beta^2-\Omega^2){\bf I}+\kappa{\bf K} +\nu{\bf N}.
\ee
Applying the Leverrier-Barnett algorithm \cite{Ba89} to the matrix polynomial $\bf L$
we find the characteristic polynomial of the system
\ba{i3}
P(\lambda)&=&\lambda^4+\delta{\rm tr}{\bf D}\lambda^3+
(2(\beta^2+\Omega^2)+\delta^2\det{\bf D}+\kappa{\rm tr}{\bf K})\lambda^2\nn\\
&+&
(4\Omega\nu+\delta(\beta^2-\Omega^2){\rm tr}{\bf D}+\delta\kappa({\rm tr}{\bf K}{\rm tr}{\bf D}-{\rm tr}{\bf KD}))\lambda\nn\\&+&\kappa^2\det{\bf K}+
\kappa{\rm tr}{\bf K}(\beta^2-\Omega^2)+(\beta^2-\Omega^2)^2+\nu^2.
\ea
In the absence of dissipative, external potential, and non-conservative positional forces $(\delta=\kappa=\nu=0)$
the characteristic polynomial \rf{i3} corresponding to the operator ${\bf L}_0(\Omega)={\bf I}\lambda^2 +
2\lambda\Omega{\bf G} + (\beta^2-\Omega^2){\bf I}$, which belongs to the class of
matrix polynomials considered, e.g., in \cite{HL01}, has four purely imaginary roots
\ba{i4}
 \lambda_p^+&=& i\beta + i\Omega,\quad
 \lambda_n^+=-i\beta + i\Omega\nn\\
 \lambda_p^-&=& i\beta - i\Omega,\quad
 \lambda_n^-=-i\beta - i\Omega.
\ea
In the plane $(\Omega,{\rm Im}\lambda)$ these roots considered as functions of $\Omega$ originate
a collection of straight lines intersecting with each other (Fig.~\ref{fig1}) -- \textit{the spectral mesh} \cite{GK06}.
Two nodes of the mesh at $\Omega=0$ correspond
to the double semi-simple eigenvalues $\lambda=\pm i \beta$. At the other two nodes at $\Omega=\pm \Omega_d$
there exist double semi-simple eigenvalues $\lambda=0$.
The range $|\Omega|<\Omega_d=\beta$ is called \textit{subcritical} for the
gyroscopic parameter $\Omega$ since the zero eigenvalue corresponds to the divergence boundary \cite{HL01}.

The double semi-simple eigenvalue $i\beta$ at $\Omega=\Omega_0=0$ has two linearly-independent eigenvectors ${\bf u}_1$ and ${\bf u}_2$
\be{i5}
{\bf u}_1=\frac{1}{\sqrt{2\beta}}\left(
            \begin{array}{c}
              0 \\
              1 \\
            \end{array}
          \right),\quad
{\bf u}_2=\frac{1}{\sqrt{2\beta}}\left(
            \begin{array}{c}
              1 \\
              0 \\
            \end{array}
          \right).
\ee Any linear combination of these vectors is an eigenvector too.
The eigenvectors are orthogonal ${\bf u}_2^T{\bf u}_1={\bf
u}_1^T{\bf u}_2=0$ and satisfy the normalization condition ${\bf
u}_1^T{\bf u}_1={\bf u}_2^T{\bf u}_2=(2\beta)^{-1}$.

Under perturbation of the gyroscopic parameter $\Omega=\Omega_0+\Delta \Omega$, the double eigenvalue $i\beta$
into two simple ones bifurcates. The asymptotic formula for the perturbed eigenvalues is \cite{KMS05}
\be{i6}
\lambda_p^{\pm}=i\beta +i\Delta\Omega\frac{{f}_{11}+{f}_{22}}{2}\pm i\Delta \Omega \sqrt{\frac{(f_{11}-f_{22})^2}{4}+f_{12}f_{21}}
\ee
where the quantities $f_{ij}$ are
\be{i7}
f_{ij}=\left.{\bf u}_j^T\frac{\partial {\bf L}_0(\Omega)}{\partial \Omega} {\bf u}_i\right|_{\Omega=0,\lambda=i\beta}=
2i\beta
{\bf u}_j^T{\bf G} {\bf u}_i.
\ee
Since $\bf G$ is a skew-symmetric matrix, we find
\be{i8}
f_{11}=f_{22}=0,\quad
f_{12}=-f_{21}=i,
\ee
and $\lambda_p^{\pm}=i\beta \pm i \Delta \Omega=i\beta \pm i \Omega$ in agreement
with \rf{i4}.

\section{Conservative, dissipative, and non-conservative deformation of the spectral mesh}

Following the approach of \cite{KMS05}, we consider a perturbation of the gyroscopic system
${\bf L}_0(\Omega)+\Delta {\bf L}(\Omega)$ by
dissipative, potential, and non-conservative positional forces breaking the symmetries of the
initial system. Assuming that the size of the perturbation
$\Delta {\bf L}(\Omega)=\delta\lambda{\bf D}+\kappa{\bf K}+\nu {\bf N}\sim \varepsilon$ is small,
where $\varepsilon=\| \Delta{\bf L}(0) \|$ is the Frobenius norm of the perturbation at $\Omega=0$ corresponding to the
double eigenvalue $i\beta$, the behavior of the perturbed eigenvalues for small $\Omega$ and small $\varepsilon$
is described by the following asymptotic formula \cite{KMS05}
\ba{i9}
\lambda_p^{\pm}&=&i\beta + i\Omega\frac{f_{11}+f_{22}}{2}+i\frac{\epsilon_{11}+\epsilon_{22}}{2}\nn\\
&\pm&i\sqrt{\frac{(\Omega(f_{11}-f_{22})+\epsilon_{11}-\epsilon_{22})^2}{4}+
(\Omega f_{12}+\epsilon_{12})(\Omega f_{21}+\epsilon_{21})},
\ea
where the coefficients $f_{ij}$ are given by the expressions \rf{i7} and $\epsilon_{ij}$ are small complex numbers
of order $\varepsilon$
\be{i10}
\epsilon_{ij}={\bf u}_j^T\Delta{\bf L}(0){\bf u}_i=
i\beta\delta{\bf u}_j^T{\bf D}{\bf u}_i+\kappa{\bf u}_j^T{\bf K}{\bf u}_i+\nu {\bf u}_j^T{\bf N}{\bf u}_i.
\ee
With the vectors \rf{i5} we find
\ba{i11}
\epsilon_{11}&=&i\frac{\delta}{2} d_{22}+ \frac{\kappa}{2\beta}k_{22},\quad
\epsilon_{12}=i\frac{\delta}{2} d_{12}+ \frac{\kappa}{2\beta}k_{12}+\frac{\nu}{2\beta}   \nn\\
\epsilon_{22}&=&i\frac{\delta}{2} d_{11}+ \frac{\kappa}{2\beta}k_{11},\quad
\epsilon_{21}=i\frac{\delta}{2} d_{12}+ \frac{\kappa}{2\beta}k_{12}-\frac{\nu}{2\beta},
\ea
where $d_{ij}$ and $k_{ij}$ denote the entries of the matrices $\bf D$ and $\bf K$. Substituting the quantities \rf{i8} and \rf{i11} into
equation \rf{i9} we obtain
\be{i12}
\lambda_p^{\pm}=i\beta+i \frac{\rho_1+\rho_2}{4\beta}\kappa - \frac{\mu_1+\mu_2}{4}\delta \pm \sqrt{c},
\ee
where
\be{i13}
c=\left(\frac{\mu_1-\mu_2}{4}\right)^2\delta^2-\left(\frac{\rho_1-\rho_2}{4\beta} \right)^2\kappa^2+
\left(i\Omega +\frac{\nu}{2\beta} \right)^2-i\delta\kappa\frac{2{\rm tr}{\bf KD}-
{\rm tr}{\bf K}{\rm tr}{\bf D}}{8\beta},
\ee
and $\mu_1$, $\mu_2$ and $\rho_1$, $\rho_2$ are eigenvalues of the matrices $\bf D$ and $\bf K$, respectively,
and thus satisfy the equations
\ba{i14}
\mu^2-\mu{\rm tr}{\bf D}+\det{\bf D}&=&0, \nn\\
\rho^2-\rho{\rm tr}{\bf K}+\det{\bf K}&=&0.
\ea

Separation of real and imaginary parts in equation \rf{i12} yields the relations
\ba{i15}
4\left({\rm Re}\lambda+\frac{{\rm tr}{\bf D}}{4}\delta \right)^4-
4\left({\rm Re}\lambda+\frac{{\rm tr}{\bf D}}{4}\delta\right)^2{\rm Re}c-
({\rm Im}c)^2&=&0,\nn\\
4\left({\rm Im}\lambda-\beta-\frac{{\rm tr}{\bf K}}{4\beta}\kappa \right)^4+
4\left({\rm Im}\lambda-\beta-\frac{{\rm tr}{\bf K}}{4\beta}\kappa \right)^2{\rm Re}c-
({\rm Im}c)^2&=&0,
\ea
where
\be{i16}
{\rm Re}c=\left(\frac{\mu_1-\mu_2}{4}\right)^2\delta^2-\left(\frac{\rho_1-\rho_2}{4\beta} \right)^2\kappa^2-\Omega^2+\frac{\nu^2}{4\beta^2},\quad
{\rm Im}c=\frac{\Omega\nu}{\beta}-\delta\kappa\frac{2{\rm tr}{\bf KD}-
{\rm tr}{\bf K}{\rm tr}{\bf D}}{8\beta}.
\ee

From the equations \rf{i15} explicit expressions follow for the real and imaginary parts
of the eigenvalues originated after the
splitting of the double semi-simple eigenvalue $i\beta$
\ba{i17}
{\rm Re}\lambda&=&-\frac{\mu_1+\mu_2}{4}\delta\pm
\sqrt{\frac{{\rm Re}c+{\sqrt{({\rm Re}c)^2+({\rm Im}c)^2}}}{2}},\nn\\
{\rm Im}\lambda&=&\beta+\frac{\rho_1+\rho_2}{4\beta}\kappa\pm
\sqrt{\frac{-{\rm Re}c+{\sqrt{({\rm Re}c)^2+({\rm Im}c)^2}}}{2}}.
\ea

The formulas \rf{i12}-\rf{i17} describe splitting of the double eigenvalues at the nodes of the spectral mesh
due to variation of parameters, including those corresponding to dissipative and non-conservative positional
forces. Such splitting obviously leads to the deformation of the mesh, in particular, to the \textit{veering}
and \textit{merging} of eigenvalue branches. Although the veering phenomenon in the systems with gyroscopic
coupling was studied both numerically and analytically, e.g in \cite{IS73,L74,Sh84,PM86,YH95,TH99,VV05}, the explicit
expressions \rf{i12}-\rf{i17} for the splitting of the double eigenvalues due to action of forces of all types were not
previously derived. Note that the approach used in our paper differs from that of the cited works because it
is based on the perturbation theory of multiple eigenvalues \cite{KS04,KMS05,KS05}. The advantage of our approach is the
description of the spectrum
of the perturbed system by means of only the derivatives of the operator with respect to parameters and
the eigenvectors of the multiple eigenvalue calculated directly at a node of the spectral mesh.

\subsection{Conservative deformation of the spectral mesh}

We first study the influence of a conservative perturbation with the matrix $\bf K$ on the spectral mesh
of the gyroscopic system. Substitution of $\delta=\nu=0$ into the formulas \rf{i15} and \rf{i16} yields
\be{c1}
\left({\rm Im}\lambda-\beta-\frac{\rho_1+\rho_2}{4\beta}\kappa\right)^2-\Omega^2=
\left(\frac{\rho_1-\rho_2}{4\beta} \right)^2\kappa^2,\quad {\rm Re}\lambda=0.
\ee

    \begin{figure}
    \begin{center}
    \includegraphics[angle=0, width=0.95\textwidth]{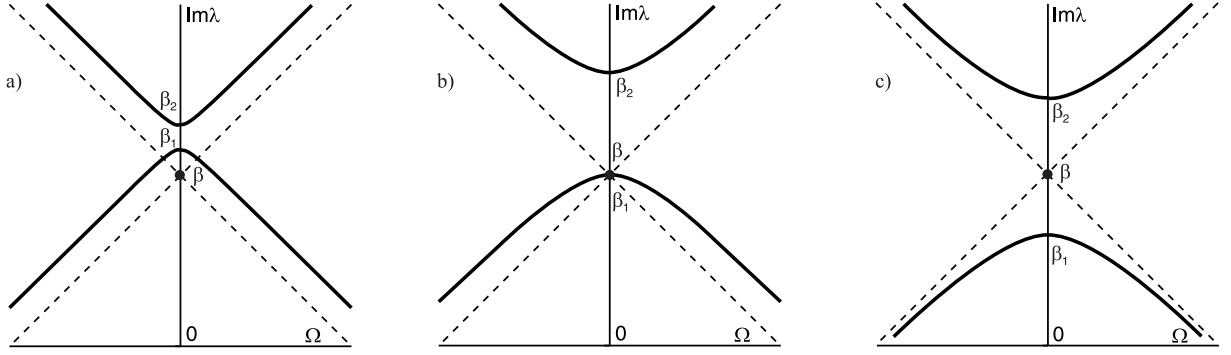}
    \end{center}
    \caption{Conservative deformation of the spectral mesh ($\kappa>0$):
    ${\bf K}$ is positive-definite (a);
    ${\bf K}$ is positive semi-definite (b);
    ${\bf K}$ is indefinite (c). }
    \label{fig1c}
    \end{figure}

When $\kappa \ne 0$, equation \rf{c1} describes a hyperbola in the plane ${\rm Im}\lambda$ versus $\Omega$.
Location of the branches of the hyperbola with respect to the lines of the unperturbed
spectral mesh ${\rm Im}\lambda=\beta \pm\Omega$ depends on the definiteness of the matrix $\bf K$,
see Fig.~\ref{fig1c}. Indeed, according to equation \rf{c1} the hyperbola has the asymptotes
\be{c2}
{\rm Im}\lambda=\beta+\frac{\rho_1+\rho_2}{4\beta}\kappa \pm \Omega.
\ee
The asymptotes cross each other above the node $(0,\beta)$ of the non-deformed spectral mesh for
${\rm tr}{\bf K}>0$, exactly at the node for $\rho_1=-\rho_2$, and below the node for
${\rm tr}{\bf K}<0$. The branches of the hyperbola intersect the axis $\Omega=0$ at the points
\be{c3}
\beta_1=\beta+\frac{\rho_1}{2\beta}\kappa, \quad \beta_2=\beta+\frac{\rho_2}{2\beta}\kappa.
\ee
If the eigenvalues $\rho_{1,2}$ have the same sign, the intersection points are always located
above or below the node of the spectral mesh for ${\bf K}>0$ or ${\bf K}<0$, respectively, see
Fig.~\ref{fig1c}(a). In case when one of the eigenvalues $\rho_{1,2}$ is zero implying
semi-definiteness of the matrix $\bf K$, one of the branches of the hyperbola goes through
the node of the spectral mesh and the other crosses the axis $\Omega=0$ above the node,
if $\bf K$ is positive semi-definite, or below it, if $\bf K$ is negative semi-definite, Fig.~\ref{fig1c}(b).
If $\bf K$ is indefinite, one of the intersection points $\beta_{1,2}$ is located above the node and another one
below the node, as indicated in Fig.~\ref{fig1c}(c).

Therefore, the conservative deformation of the spectral mesh
does not shift the eigenvalues from the imaginary axis near the nodes $(0,\pm\beta)$ and thus preserves the
marginal stability. However, the deformation pattern depends on the definiteness of the perturbation
matrix $\bf K$. In particular, in the degenerate case when $\det{\bf K}=0$, one of the eigenvalue branches,
originated after the splitting of the double eigenvalue, always passes through the point corresponding to the
node of the unperturbed spectral mesh, Fig.~\ref{fig1c}(b).

\subsection{Creation and activation of the latent sources of instability by dissipation}

\begin{figure}
    \begin{center}
    \includegraphics[angle=0, width=0.95\textwidth]{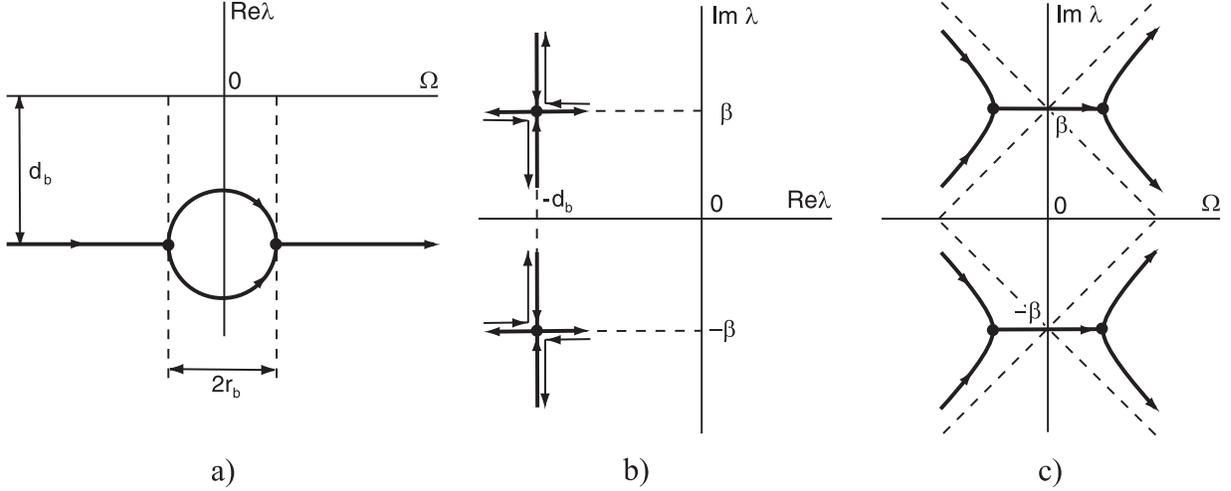}
    \end{center}
    \caption{Origination of a latent source of the subcritical flutter instability in presence of full dissipation:
    Submerged bubble of instability (a);
    coalescence of eigenvalues in the complex plane at two exceptional points (b);
    hyperbolic trajectories of imaginary parts (c). }
    \label{fig2}
    \end{figure}

\begin{figure}
    \begin{center}
    \includegraphics[angle=0, width=0.85\textwidth]{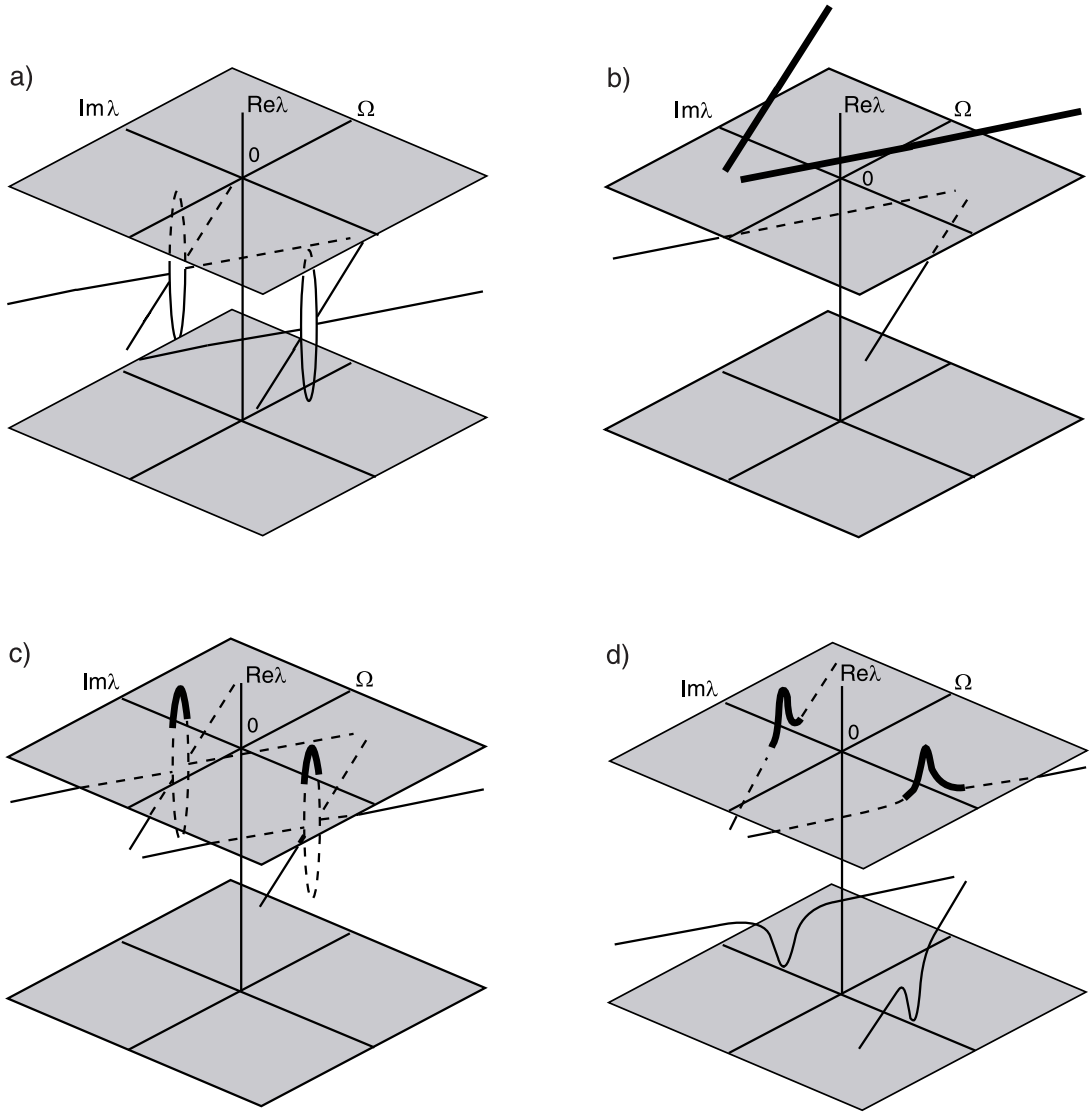}
    \end{center}
    \caption{The mechanism of subcritical flutter instability (bold lines):
    The ring (bubble) of complex eigenvalues submerged under the surface
    ${\rm Re}\lambda=0$ due to action of dissipation with $\det {\bf D} \ge 0$
    - a latent source of instability (a);
    repulsion of eigenvalue branches of the spectral mesh due to action of non-conservative positional forces (b);
    emersion of the bubble of instability due to indefinite damping with $\det {\bf D} < 0$ (c);
    collapse of the bubble of instability and immersion and emersion of its parts due to combined action of
    dissipative and non-conservative positional forces (d). }
    \label{fig3}
    \end{figure}

Now we consider the effect of dissipative forces on the stability of the gyroscopic system
and study its dependence on the properties of the matrix $\bf D$. Assuming $\nu=\kappa=0$ in expression \rf{i13}
we rewrite formula \rf{i12} in the form
\be{i18}
\lambda=i\beta - \frac{\mu_1+\mu_2}{4}\delta \pm
\sqrt{\left(\frac{\mu_1-\mu_2}{4}\right)^2\delta^2-\Omega^2}.
\ee
Since ${\rm Im}c=0$, equations \rf{i15} can be transformed into
\be{i19}
\left({\rm Re}\lambda+\frac{\mu_1+\mu_2}{4}\delta\right)^2+\Omega^2=
\frac{(\mu_1-\mu_2)^2}{16}\delta^2,\quad {\rm Im}\lambda=\beta
\ee
when
\be{i20}
\Omega^2-
\frac{(\mu_1-\mu_2)^2}{16}\delta^2<0,
\ee
and into
\be{i21}
\Omega^2-\left({\rm Im}\lambda-\beta\right)^2=
\frac{(\mu_1-\mu_2)^2}{16}\delta^2,\quad
{\rm Re}\lambda=-\frac{\mu_1+\mu_2}{4}\delta,
\ee
when the sign in inequality \rf{i20} is opposite.
For a given $\delta$ equation \rf{i21} defines a hyperbola in the plane $(\Omega,{\rm Im}\lambda)$, while
\rf{i19} is the equation of a circle in the plane $(\Omega,{\rm Re}\lambda)$, as shown in Fig.~\ref{fig2}(a,c).
For tracking the complex eigenvalues due to change of the gyroscopic parameter $\Omega$, it is convenient to
consider the eigenvalue branches in the three-dimensional space $(\Omega, {\rm Im}\lambda, {\rm Re}\lambda)$.
In this space the circle belongs to the plane ${\rm Im}\lambda=\beta$ and the hyperbola lies in the
plane ${\rm Re}\lambda=-\delta(\mu_1+\mu_2)/4$, see Fig.~\ref{fig3}(a,c). Below we show that the circle
of the complex eigenvalues -- \textit{the bubble of instability} \cite{KS02a} -- plays the crucial role in the development
and localization of the subcritical flutter instability and that its properties depend on whether the
matrix $\bf D$ is definite or indefinite.

\subsubsection{Full dissipation and pervasive damping: A latent state of the bubble of instability}

From the formulas \rf{i19} and \rf{i21} we see that the basic geometric characteristics of the trajectories
of eigenvalues of the dissipatively perturbed gyroscopic system are defined by the eigenvalues $\mu_1$
and $\mu_2$ of the matrix $\bf D$. For example, the radius of the bubble of instability $r_b$ and the distance $d_b$
of its center from the plane ${\rm Re} \lambda=0$ are
\be{i22}
r_b=\frac{|(\mu_1-\mu_2)\delta|}{4},\quad d_b=\frac{|(\mu_1+\mu_2)\delta|}{4}.
\ee

As a consequence, the bubble does not intersect the plane ${\rm Re} \lambda=0$ under the condition $d_b\ge r_b$,
which is equivalent to the inequality
\be{i23}
\mu_1\mu_2=\det{\bf D}\ge 0.
\ee
Moreover, the bubble of instability is "submerged" under the surface ${\rm Re}\lambda=0$ in the space
$(\Omega, {\rm Im}\lambda, {\rm Re}\lambda)$, if
\be{i24}
\delta(\mu_1+\mu_2)=\delta{\rm tr}{\bf D}>0.
\ee

Since the inequalities \rf{i23} and \rf{i24} imply positive semi-definiteness of the matrix $\delta\bf D$,
we conclude that the role of full dissipation or pervasive damping is to deform the spectral mesh in such a way that
the double semi-simple eigenvalue is inflated to the bubble of complex eigenvalues \rf{i19}
connected with the two branches of the hyperbola \rf{i21} at the points
\be{i25}
{\rm Im}\lambda=\beta,\quad
{\rm Re}\lambda=-\delta (\mu_1+\mu_2)/4,\quad
\Omega=\pm \delta (\mu_1-\mu_2)/4,
\ee
and to plunge all the eigenvalue curves into the region ${\rm Re}\lambda\le 0$.
The eigenvalues at the points \rf{i25} are double and have a Jordan chain of generalized eigenvectors of order 2.
In the complex plane the eigenvalues of the
perturbed system move with the variation of $\Omega$ along the lines ${\rm Re}\lambda=-d_b$
until they meet at the points \rf{i25} and then split in the orthogonal direction; however, they never cross the imaginary axis,
see Fig.~\ref{fig2}(b).
Note that a similar process of unfolding the semi-simple eigenvalue (diabolical point) into two double
eigenvalues with the Jordan blocks (exceptional points) for Hermitian matrices under arbitrary complex perturbation
and its application to crystal optics have been described in \cite{BD03,KMS05}.

The bubble of instability has two remarkable properties important for the explanation of the phenomenon of squeal.
First, there can exist perturbations causing its growth and emersion above the surface ${\rm Re}\lambda=0$ (flutter);
second, for small $\delta$ the instability is localized in the vicinity of the frequency ${\rm Im}\lambda=\beta$ and
the value of the gyroscopic parameter $\Omega=0$. The narrow frequency band is a characteristic property of squeal.
We conclude that the bubble is a latent source of subcritical flutter instability localized
in a narrow range of change of the gyroscopic parameter with $|\Omega|<\Omega_d$ at the frequency corresponding to the
double eigenvalue of the non-rotating system.

\subsubsection{Indefinite damping: An active state of the bubble of instability}

As it is seen from equations \rf{i22}, the radius of the bubble of instability is greater then the depth
of its submersion under the surface ${\rm Re}\lambda=0$ only if the eigenvalues $\mu_1$ and $\mu_2$
of the damping matrix have different signs, i.e. if \textit{the damping is indefinite}. The damping
with the indefinite matrix appears for example in the systems with frictional contact when the friction
coefficient is decreasing with relative sliding velocity as well as in the problems of propagation of waves
in dissipative media \cite{Fr97,PP02,KOP03}. Indefinite damping is known to be a destabilizing factor
\cite{KM97,Ko07,Ki07}. In our case it leads to the emersion of the bubble of instability meaning that
the eigenvalues of the bubble have positive real parts in the range
$\Omega^2<\Omega_{cr}^2$, where
\be{i26}
\Omega_{cr}=\frac{\delta}{2}\sqrt{-\det{\bf D}}
\ee
is found from the equation \rf{i19} after assuming there ${\rm Re}\lambda=0$. The right side of the formula \rf{i26}
is real only for $\det{\bf D}<0$, i.e. for the indefinite matrix $\bf D$.

    \begin{figure}
    \begin{center}
    \includegraphics[angle=0, width=0.85\textwidth]{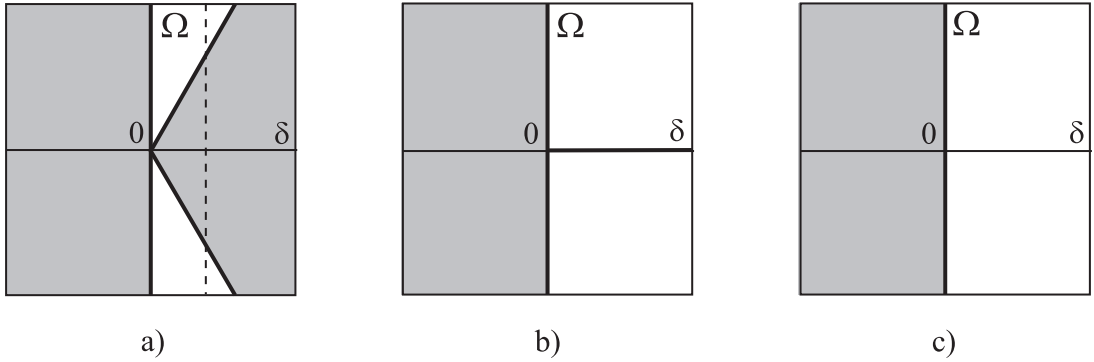}
    \end{center}
    \caption{Approximation based on the first-order perturbation theory of double eigenvalues
    to the domain of asymptotic stability (white) and its boundary (bold lines)
    for the dissipatively perturbed gyroscopic system (1) when $\nu=\kappa=0$, ${\rm tr}{\bf D}>0$ and $\det{\bf D}<0$ (a);
    $\det{\bf D}=0$ (b); $\det{\bf D}>0$ (c).}
    \label{fig3a}
    \end{figure}

We see that the domain of asymptotic stability is defined by the constraints $\delta{\rm tr}{\bf D}>0$ and
$\Omega^2>\Omega_{cr}^2$. In the plane of the parameters $\delta$ and $\Omega$ for $\det{\bf D}<0$ it has
a form shown in Fig.~\ref{fig3a}(a). Due to the singularity existing at the origin, an unstable system with
indefinite damping can be stabilized by sufficiently strong gyroscopic forces, as shown by the dashed line in
Fig.~\ref{fig3a}(a). With the increase of $\det{\bf D}$ the stability domain is getting wider and for $\det{\bf D}>0$
it is defined by the condition $\delta{\rm tr}{\bf D}>0$, Fig.~\ref{fig3a}(c).
At $\det{\bf D}=0$ the line $\Omega=0$ does not belong to the domain of asymptotic stability, Fig.~\ref{fig3a}(b).

Therefore, changing the damping matrix $\delta \bf D$ from positive definite to indefinite we trigger
the state of the bubble of instability from latent $({\rm Re}\lambda<0)$ to active $({\rm Re}\lambda>0)$,
see Fig.~\ref{fig3}(a,c). Since for small $\delta$ we have $\Omega_{cr}<\Omega_d$, the flutter instability
is localized in the vicinity of $\Omega=0$. As we show below, the non-conservative positional forces play the
similar role.

\subsection{Activation of the bubble of instability by non-conservative positional forces}

In the absence of dissipation non-conservative positional forces destroy the marginal stability
of gyroscopic systems \cite{La75,Ka75}. One can easily check that this property is valid for system (1)
by assuming $\delta=\kappa=0$ in the formulas \rf{i12} and \rf{i13}, which yield
\be{i27}
\lambda_p^{\pm}= i\beta \pm i\Omega \pm \frac{\nu}{2\beta},\quad
\lambda_n^{\pm}=-i\beta \pm i\Omega \mp \frac{\nu}{2\beta}.
\ee

Equations \rf{i27} show that the eigenvalues of the branches $i\beta+i\Omega$ and $-i\beta-i\Omega$
of the spectral mesh get positive real parts due to perturbation by the non-conservative positional forces.
The eigenvalues of the other two branches are shifted to the left from the imaginary axis, see Fig.~\ref{fig3}(b).
Note that the real part of the perturbed eigenvalue is proportional to $1/\beta$, which means that the
destabilizing effect of non-conservative positional forces is less pronounced for higher frequencies.
This is important in the study of systems with many degrees of freedom.
Thus, the instability induced by the non-conservative forces only is not localized
near the nodes of the spectral mesh, in contrast to the effect of indefinite damping.
We show now that in combination
with the dissipative forces, both definite and indefinite, the non-conservative forces can create
subcritical flutter instability in the vicinity of diabolical points.

    \begin{figure}
    \begin{center}
    \includegraphics[angle=0, width=0.95\textwidth]{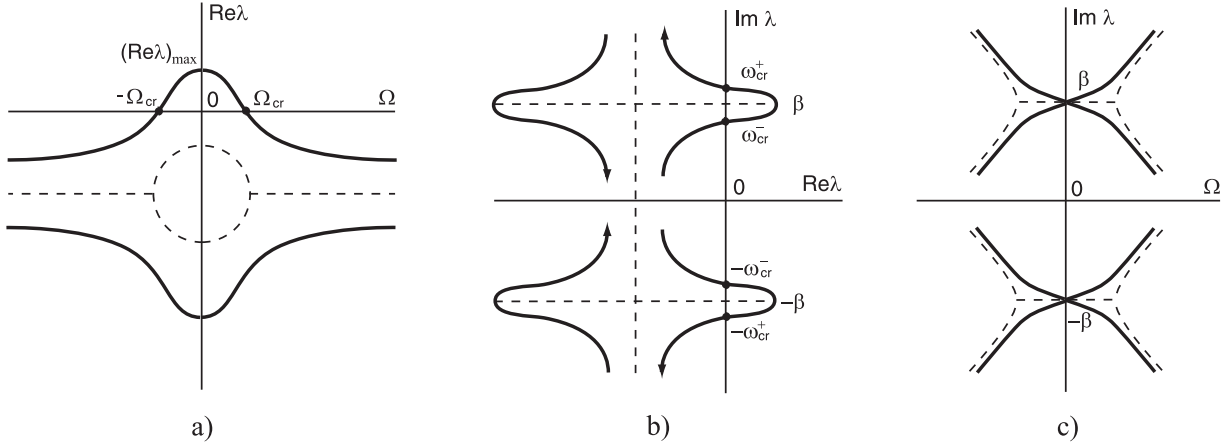}
    \end{center}
    \caption{Subcritical flutter instability due to combined action of dissipative and non-conservative positional forces:
    Collapse and emersion of the bubble of instability (a);
    excursions of eigenvalues to the right side of the complex plane when
    $\Omega$ goes from negative values to positive (b);
    crossing of imaginary parts (c). }
    \label{fig4}
    \end{figure}

Recall that the real and imaginary parts of the eigenvalues originated after the spitting of the double
eigenvalues due to combined action of dissipative and non-conservative positional forces are given
by the expressions \rf{i15} and \rf{i17}, where we assume $\kappa=0$. Multiplying equations \rf{i17} we find that the trajectories
of the eigenvalues in the complex plane are described by the formula
\be{i28}
\left({\rm Re}\lambda +\frac{{\rm tr}{\bf D}}{4}\delta \right)\left({\rm Im}\lambda-\beta \right)=
\frac{\Omega \nu}{2\beta}.
\ee
When $\nu=0$ and $\delta\ne0$ is given, the eigenvalues move  with the variation of the gyroscopic parameter $\Omega$
along the lines ${\rm Re}\lambda=-{\rm tr}{\bf D}/4$ and ${\rm Im}\lambda=\beta$ and merge at the points
\rf{i25}, see Fig.~\ref{fig2}(b).

Non-conservative positional forces with $\nu\ne 0$ destroy the merging of modes. As a consequence, the eigenvalues move
along the separated trajectories. According to relations \rf{i27} the eigenvalues with $|{\rm Im}\lambda|$
increasing due to an increase in $|\Omega|$ move closer to the imaginary axis then the others,
as shown in Fig~\ref{fig4}(b). Therefore, in the space $(\Omega,{\rm Im}\lambda,{\rm Re}\lambda)$
the action of the non-conservative positional forces separates the bubble of instability and the
adjacent hyperbolic eigenvalue branches into two non-intersecting curves, see Fig~\ref{fig3}(d).
It is remarkable that the form of each of the new eigenvalue curves carries the memory about the original
bubble of instability. Due to this spectral peculiarity the real parts of the eigenvalues can be
positive for the values of the gyroscopic parameter localized near $\Omega=0$ in the range
$\Omega^2<\Omega_{cr}^2$. Taking into account that ${\rm Re}\lambda=0$
at the critical values of the gyroscopic parameter we find $\Omega_{cr}$ from the equations \rf{i17}
\be{i29}
\Omega_{cr}=\delta\frac{{\rm tr}{\bf D}}{4}\sqrt{-\frac{
\nu^2-\delta^2\beta^2\det{\bf D}}{
\nu^2-\delta^2\beta^2({\rm tr}{\bf D}/2)^2}}.
\ee
Additionally, it follows from \rf{i17} that the eigenfrequencies of the unstable modes from the interval
$\Omega^2<\Omega_{cr}^2$ are localized near the frequency of the double semi-simple eigenvalue
at the node of the undeformed spectral mesh: $\omega_{cr}^-<\omega<\omega_{cr}^+$
\be{i30}
\omega_{cr}^{\pm}=
\beta\pm\frac{\nu}{2\beta}\sqrt{-\frac{
\nu^2-\delta^2\beta^2\det{\bf D}}{
\nu^2-\delta^2\beta^2({\rm tr}{\bf D}/2)^2}}.
\ee

When the radicand in formulas \rf{i29} and \rf{i30} is real, the eigenvalues move in the complex plane
making the excursion to its right side, as shown in Fig.~\ref{fig4}(b). As it follows from \rf{i29},
in presence of non-conservative positional forces such excursions behind the stability boundary
are possible for the eigenvalues, even when dissipation is full $(\det{\bf D}>0)$. One can say that
similarly to the indefinite damping \textit{the non-conservative positional forces activate the latent sources of flutter instability created by
the full dissipation}.

\begin{figure}
    \begin{center}
    \includegraphics[angle=0, width=0.95\textwidth]{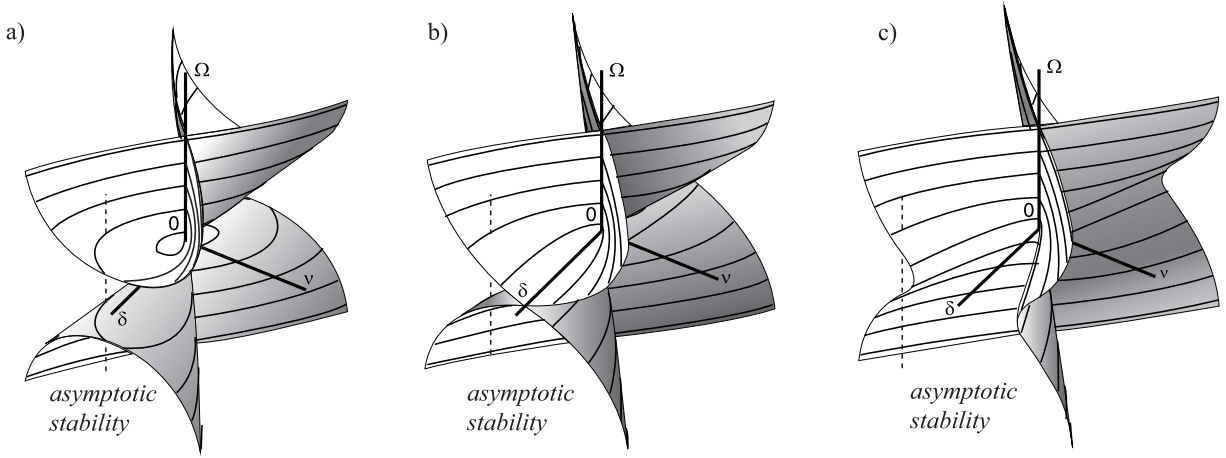}
    \end{center}
    \caption{Domains of asymptotic stability in the space $(\delta,\nu,\Omega)$ for different types
    of damping: Indefinite damping $\det{\bf D}<0$ (a); semi-definite (pervasive) damping $\det{\bf D}=0$ (b);
    full dissipation $\det{\bf D}>0$ (c).}
    \label{fig5}
    \end{figure}

The equation \rf{i29} describes the surface in the space of the parameters $\delta$, $\nu$, and $\Omega$,
which is an approximation to the stability boundary separating
the domains of asymptotic stability and flutter. For better understanding the shape of this surface
we extract the parameter $\nu$ in \rf{i29}
\be{i31}
\nu=
\pm\delta\beta{\rm tr}{\bf D}\sqrt{
\frac{\delta^2\det{\bf D}+4\Omega^2}{\delta^2({\rm tr}{\bf D})^2+16\Omega^2}}.
\ee
If $\det{\bf D}\ge 0$ and $\Omega$ is fixed, the formula \rf{i31} describes two independent curves in the plane $(\delta,\nu)$ intersecting
with each other at the origin along the straight lines given by the expression
\be{i32}
\nu=\pm\frac{\beta{\rm tr}{\bf D}}{2}\delta.
\ee
However, in case when damping is indefinite and $\det{\bf D}<0$, the radical in the formula \rf{i31}
is real only for $\delta^2<-4\Omega^2/\det{\bf D}$ meaning that \rf{i31} describes two branches of a closed loop
in the plane of the parameters $\delta$ and $\nu$. The loop is smooth at every point except at the origin, where it is
self-intersecting with the tangents given by the expression \rf{i32}. Therefore, for a given $\Omega$ this curve looks
like figure "8". When $\Omega$ goes to zero, the size of the self-intersecting curve tends to zero too. We conclude
that in case $\det{\bf D}<0$ the shape of the surface described by equation \rf{i29} or \rf{i31} is a cone with the "8"-shaped loop in
a cross-section, see Fig.~\ref{fig5}(a). Due to self-intersections the cone consists of four pockets.
The asymptotic stability domain is inside the two of them, selected by the inequality \rf{i24}, as shown in
Fig.~\ref{fig5}(a). The singularity of the stability domain at the origin is the degeneration
of a more general configuration described first in \cite{Ki07}.

The domain of asymptotic stability bifurcates when $\det{\bf D}$ changes from negative to positive values.
This process is shown in Fig.~\ref{fig5}. Indeed, according to formula \rf{i31} with increasing $\det{\bf D}<0$
two pockets of the domain of asymptotic stability move towards each other until they have a common tangent line
$\nu=0$ at $\det{\bf D}=0$, see Fig.~\ref{fig5}(b). When $\det{\bf D}$ is positive,
this temporarily glued configuration unfolds to the unique domain of asymptotic stability bounded by the
two surfaces intersecting along the $\Omega$-axis, as shown in Fig.~\ref{fig5}(c).

In Fig.~\ref{fig5}(a) we see that in case of indefinite damping there always exists an instability gap due to
the singularity at the origin. Starting in the flutter domain at $\Omega=0$ for any combination of the parameters
$\delta$ and $\nu$ one can reach the domain of asymptotic stability at higher values of $|\Omega|$
(gyroscopic stabilization), as shown in Fig.~\ref{fig5}(a) by the dashed line.
The gap is responsible for the subcritical flutter
instability localized in the vicinity of the node of the spectral mesh of the unperturbed gyroscopic system.
When $\det{\bf D}=0$, the gap vanishes in the direction $\nu=0$. In case of full dissipation $(\det{\bf D}>0)$
the singularity at the origin unfolds. However, the memory about it is preserved in
the two instability gaps located in the folds of the stability boundary with the locally strong curvature,
Fig.~\ref{fig5}(c).

We see that in case of full dissipation for some combinations of the parameters $\delta$ and $\nu$
the system is asymptotically stable at any $\Omega$. There exist, however, the values of $\delta$ and $\nu$
for which one can penetrate the fold of the stability boundary with the change of $\Omega$, as shown in
Fig.~\ref{fig5}(c) by the dashed line. For such $\delta$ and $\nu$ the flutter instability is localized in the vicinity of $\Omega=0$.
It is remarkable that in presence of non-conservative positional forces the system with full dissipation
can suffer from the subcritical flutter localized near the nodes of the spectral mesh. A good illustration
of this fact is the formula for the maximal real part of the unstable eigenvalue attained at $\Omega=0$ (see
Fig.~\ref{fig4}(a))
\be{i33}
({\rm Re}\lambda)_{\rm max}=
-\frac{\mu_1+\mu_2}{4}\delta+
\sqrt{\left(\frac{\mu_1-\mu_2}{4}\right)^2\delta^2+\frac{\nu^2}{4\beta^2}}.
\ee

From our previous considerations it follows that the phenomenon of the local subcritical flutter
instability is controlled by the eigenvalues of the matrix $\bf D$. When both of them are positive,
the folds of the stability boundary are more pronounced if one of the eigenvalues is close to zero.
If one of the eigenvalues is negative and the other is positive, the local subcritical flutter
instability is possible for any combination of $\delta$ and $\nu$ including the case when the
non-conservative positional forces are absent $(\nu=0)$.


Even if the structure of the damping matrix $\bf D$ is unknown, we realize that the main role of dissipation of any kind
is the creation of the bubble of instability. It is submerged below the surface ${\rm Re}\lambda=0$ in the
space $(\Omega,{\rm Im}\lambda,{\rm Re}\lambda)$ in case of full dissipation and partially lies in the domain
${\rm Re}\lambda>0$ when damping is indefinite. Non-conservative positional forces destroy the bubble
of instability into two branches and shift one of them to the region of positive real parts even in case
of full dissipation. Since the branch remembers the existence of the bubble, the instability is developing
locally near the nodes of the spectral mesh.

\textit{Therefore, the instability mechanism behind the squeal is the emersion (or activation) due to indefinite damping
and non-conservative positional forces of the bubbles of instability created by the full dissipation in the
vicinity of the nodes of the spectral mesh}.

\section{Example. A rotating circular string constrained by a stationary load system}

    \begin{figure}
    \begin{center}
    \includegraphics[angle=0, width=0.95
    \textwidth]{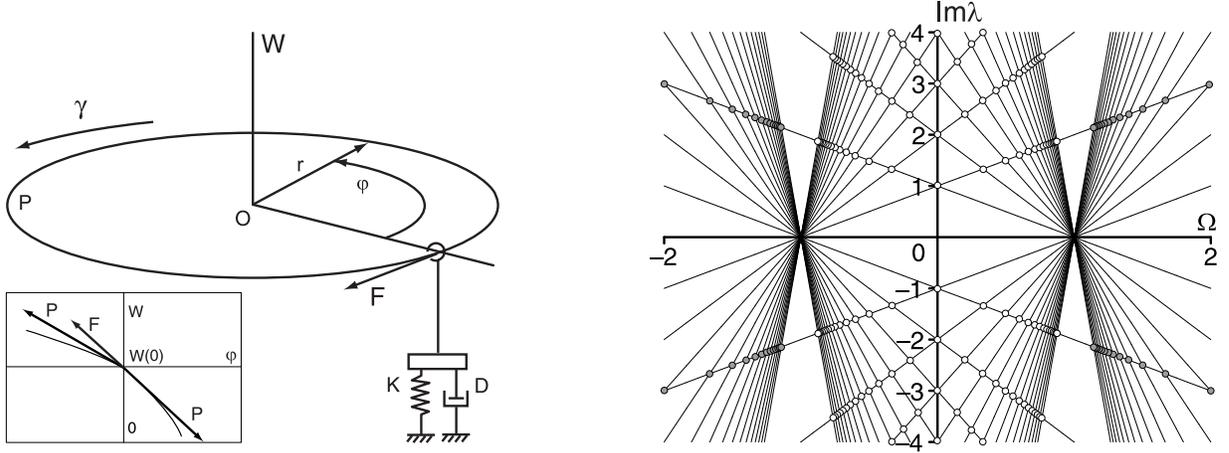}
    \end{center}
    \caption{A rotating circular string and its "keyboard" constituted by the nodes (marked by
    white and grey) of the spectral mesh (only 30 modes are shown).}
    \label{fig7a}
    \end{figure}

Following Yang and Hutton \cite{YH95}, we consider a rotating circular string of displacement
$W(\varphi, \tau)$, constrained at $\varphi=0$ by a stationary load system consisting of a spring,
a damper, and a massless eyelet generating a constant frictional follower force $F$ \cite{Mo98} on the
string, as shown in Fig.~\ref{fig7a}.
The parameters $r$ and $\rho$ are the radius and mass per unit length of the string.
The rotational speed of the string is $\gamma$. The following assumptions are adopted in developing
the governing equation of the problem: the circumferential tension $P$ in the string is constant;
the stiffness of the spring supporting the string is $K$ and the damping coefficient of the viscous damper is $D$;
the velocity of the string in the $\varphi$ direction has constant value $\gamma r$ \cite{YH95}.

Introducing the following non-dimensional variables and parameters
\be{s1}
t=\frac{\tau}{r}\sqrt{\frac{P}{\rho}},\quad w=\frac{W}{r},\quad \Omega=\gamma r\sqrt{\frac{\rho}{P}},
\quad k=\frac{Kr}{P},\quad \mu=\frac{F}{P}, \quad d=\frac{D}{\sqrt{\rho P}},
\ee
we arrive at the non-dimensional governing equation and boundary conditions \cite{YH95}:
\be{s2}
w_{tt}+2\Omega w_{t\varphi}-(1-\Omega^2)w_{\varphi \varphi}=0,
\ee
\be{s3}
w(0,t)-w(2\pi,t)=0,~~(1-\Omega^2)[w_{\varphi}(2\pi,t)-w_{\varphi}(0,t)]+kw(0,t)+dw_t(0,t)+\mu w_{\varphi}(0,t)=0.
\ee
The boundary conditions \rf{s3} reflect the continuity of the string displacement and the discontinuity of
its slope following from the force balance at the eyelet in the assumption of smallness of the norm of
$w$ and $w_{\varphi}$ with respect to $r$. The inclusion in Fig.~\ref{fig7a}
shows the mutual configuration of the vectors of the frictional follower force $F$ and of the circumferential
tension $P$.

Separation of time by the substitution $w(\varphi, t)=u(\varphi)\exp(\lambda t)$ yields the boundary value problem
\be{s4}
Lu=\lambda^2 u+ 2\Omega\lambda u'-(1-\Omega^2)u''=0,
\ee
\be{s5}
u(0)-u(2\pi)=0,\quad u'(0)-u'(2\pi)=\frac{\lambda d+k}{1-\Omega^2}u(0)+\frac{\mu}{1-\Omega^2}u'(0),
\ee
where prime denotes partial differentiation with respect to $\varphi$.
The non-self-adjoint boundary eigenvalue problem depends on four parameters $\Omega$, $d$, $k$, and $\mu$
expressing the speed of rotation, and damping, stiffness, and friction coefficients of the eyelet,
respectively.

Taking the scalar product $(Lu,v)=\int_0^{2\pi}\bar v L u d\varphi$,
where the bar over a symbol denotes complex conjugate, then integrating it by parts
and employing the boundary conditions \rf{s5} we arrive at the boundary value problem adjoint
to \rf{s4} and \rf{s5}
\be{s6}
L^*v=\bar{\lambda}^2 v- 2\Omega\bar{\lambda}v'-(1-\Omega^2)v''=0,
\ee
\be{s7}
v(0)-v(2\pi)=-\frac{\mu}{1-\Omega^2}\, v(2\pi),\quad
v'(0)-v'(2\pi)=
\frac{\bar\lambda d+k}{1-\Omega^2}v(2\pi)+
\frac{2\Omega\mu}{(1-\Omega^2)^2}v(2\pi).
\ee

Let us first consider the string without constraints ($d=k=\mu=0$). Then, the system is gyroscopic
so that the eigenfunctions of the adjoint boundary value problems corresponding to a purely imaginary
eigenvalue $\lambda$ coincide, i.e. $v = u$. Assuming the solution of the equation \rf{s4} in the form
$u=C_1 \exp\left(\frac{\lambda}{1-\Omega} \varphi\right) + C_2 \exp\left(\frac{-\lambda}{1+\Omega} \varphi\right) $
and substituting it into the boundary conditions \rf{s5} we obtain the characteristic equation
\be{s8}
8\lambda\sin\frac{\pi\lambda}{i(1-\Omega)}\sin\frac{\pi\lambda}{i(1+\Omega)}
\frac{{\rm e}^\frac{-2\pi\lambda\Omega}{\Omega^2-1}
}{\Omega^2-1}=0.
\ee
The roots of the equation \rf{s8} are
\be{s9}
\lambda_n^+=in(1+\Omega),\quad
\lambda_n^-=in(1-\Omega),
\ee
where $n\in \mathbb{Z}$. They are the eigenvalues of the boundary eigenvalue problem \rf{s4}, \rf{s5}
with the eigenfunctions
\be{s10}
u_n^+=\cos(n \varphi)-i\sin(n \varphi),\quad u_n^-=\cos(n \varphi)+i\sin(n \varphi),
\ee
respectively. The eigenvalues are purely imaginary and form a mesh of lines intersecting
with each other in the plane ${\rm Im}\lambda$ versus $\Omega$,
as shown in Fig.~\ref{fig7a}.

The eigenvalues \rf{s9} are simple almost at all values of $\Omega$ excepting those corresponding
to the nodes of the spectral mesh. Indeed, two eigenvalue branches
$\lambda_n^{\varepsilon}=in(1+\varepsilon\Omega)$ and $\lambda_m^{\delta}=im(1+\delta\Omega)$,
where $\varepsilon,\delta=\pm$, intersect each other at $\Omega=\Omega_{mn}^{\varepsilon \delta}$
\be{s11}
\Omega_{mn}^{\varepsilon\delta}=\frac{n-m}{m\delta-n\varepsilon}
\ee
and originate the double eigenvalue
\be{s12}
\lambda_{mn}^{\varepsilon\delta}=\frac{inm(\delta-\varepsilon)}{m\delta-n\varepsilon},
\ee
which has two linearly independent eigenfunctions
\be{s13}
u_n^{\varepsilon}=\cos(n \varphi)-{\varepsilon}i\sin(n \varphi),\quad u_m^{\delta}=\cos(m \varphi)-\delta i\sin(m\varphi).
\ee

The nodes \rf{s11}, \rf{s12} of the spectral mesh of the rotating circular string in the absence of
the external loading are marked by white and grey dots in Fig.~\ref{fig7a}. At $\Omega=0$ the spectrum
of the non-rotating circular string consists of the double semi-simple purely imaginary eigenvalues $in$,
$n\in\mathbb{Z}$, each of which splits into two simple purely imaginary eigenvalues due to change in the angular
velocity \cite{Sh84,PM86,YH95,TH99}. At $\Omega=\pm1$ all the eigenvalue branches cross the axis
${\rm Im}\lambda=0$, see Fig.~\ref{fig7a}. In the following we will consider the spectrum for the angular velocities
from the subcritical range when $\Omega \in (-1,1)$.

Now we consider the deformation of the spectral mesh caused by the interaction of the rotating string with
the external loading system. We proceed analogously to our investigation of a two-dimensional system and
study the splitting of the double eigenvalues at the nodes of the spectral mesh. For this purpose
we use the perturbation theory of multiple eigenvalues of non-self-adjoint differential operators developed in
\cite{KS04,KS05,GK06}.
According to this theory the perturbed eigenvalues are expressed by the following
asymptotic formula
\ba{s14}
\lambda&=&\lambda_{nm}^{\varepsilon \delta}-
\frac{f_{nn}^{\varepsilon\varepsilon}+f_{mm}^{\delta\delta}}{2}-
\frac{\epsilon_{nn}^{\varepsilon\varepsilon}+\epsilon_{mm}^{\delta\delta}}{2}\nn\\
&\pm&\sqrt{\frac{\left(f_{nn}^{\varepsilon\varepsilon}-f_{22}^{\delta\delta}+
\epsilon_{nn}^{\varepsilon\varepsilon}-\epsilon_{22}^{\delta\delta}\right)^2}{4}-
(f_{nm}^{\varepsilon\delta}+\epsilon_{nm}^{\varepsilon\delta})
(f_{mn}^{\delta\varepsilon}+\epsilon_{mn}^{\delta\varepsilon})}.
\ea
The coefficients $f_{nm}^{\varepsilon\delta}$ are defined by
\be{s15}
f_{nm}^{\varepsilon\delta}=\frac{2\lambda_{nm}^{\varepsilon\delta}\int_0^{2\pi}{u_n^{\varepsilon}}'
\bar u_m^{\delta}d\varphi+
2\Omega_{nm}^{\varepsilon\delta}\int_0^{2\pi}{u_n^{\varepsilon}}''\bar u_m^{\delta}d\varphi}
{2\sqrt{\int_0^{2\pi}(\lambda_{nm}^{\varepsilon\delta}u_n^{\varepsilon}+
\Omega_{nm}^{\varepsilon\delta}{u_n^{\varepsilon}}')\bar u_n^{\varepsilon}d\varphi
\int_0^{2\pi}(\lambda_{nm}^{\varepsilon\delta}u_m^{\delta}+
\Omega_{nm}^{\varepsilon\delta}{u_m^{\delta}}')\bar u_m^{\delta}d\varphi}}\,\Delta\Omega,
\ee
while the quantities $\epsilon_{nm}^{\varepsilon\delta}$ are
\be{s16}
\epsilon_{nm}^{\varepsilon\delta}=\frac{(d \lambda_{nm}^{\varepsilon\delta}+k){u_n^{\varepsilon}}(0)\bar u_m^{\delta}(0)+
\mu {u_n^{\varepsilon}}'(0)\bar u_m^{\delta}(0)}{2\sqrt{\int_0^{2\pi}(\lambda_{nm}^{\varepsilon\delta}u_n^{\varepsilon}+
\Omega_{nm}^{\varepsilon\delta}{u_n^{\varepsilon}}')\bar u_n^{\varepsilon}d\varphi
\int_0^{2\pi}(\lambda_{nm}^{\varepsilon\delta}u_m^{\delta}+
\Omega_{nm}^{\varepsilon\delta}{u_m^{\delta}}')\bar u_m^{\delta}d\varphi}}.
\ee
where $\Delta\Omega=\Omega-\Omega_{nm}^{\varepsilon\delta}$. Calculating the integrals in \rf{s15} and
\rf{s16} and taking into account expressions \rf{s11} and \rf{s12}, we obtain
\be{s17}
f_{nn}^{\varepsilon\varepsilon}=-i\varepsilon n \Delta\Omega,\quad
f_{nm}^{\varepsilon\delta}=f_{mn}^{\delta\varepsilon}=0,\quad
f_{mm}^{\delta\delta}=-i\delta m \Delta\Omega,
\ee
and
\ba{s18}
\epsilon_{nn}^{\varepsilon\varepsilon}=\frac{d \lambda_{nm}^{\varepsilon\delta}+k-\varepsilon in\mu}{4\pi n i}, &\quad&
\epsilon_{nm}^{\varepsilon\delta}=\frac{d \lambda_{nm}^{\varepsilon\delta}+k-\varepsilon in\mu}{4\pi i \sqrt{nm}}, \nn\\
\epsilon_{mn}^{\delta\varepsilon}=\frac{d \lambda_{nm}^{\varepsilon\delta}+k-\delta im\mu}{4\pi i \sqrt{nm}}, &\quad&
\epsilon_{mm}^{\delta\delta}=\frac{d \lambda_{nm}^{\varepsilon\delta}+k-\delta im\mu}{4\pi m i}.
\ea

Taking into account the results of our calculations \rf{s17} and \rf{s18} we find
\be{s19}
\lambda=\lambda_{nm}^{\varepsilon\delta}+i\frac{\varepsilon n +\delta m}{2}\Delta\Omega+
i\frac{n+m}{8\pi nm}(d\lambda_{nm}^{\varepsilon\delta}+k)+
\frac{\varepsilon+\delta}{8\pi}\mu\pm \sqrt{c}
\ee
where
\be{s20}
c=\left(i\frac{\varepsilon n - \delta m}{2}\Delta \Omega+
i\frac{m-n}{8\pi mn}(d\lambda_{nm}^{\varepsilon\delta}+k)+
\frac{\varepsilon-\delta}{8\pi}\mu \right)^2-
\frac{(d\lambda_{nm}^{\varepsilon\delta}+k-i\varepsilon n\mu)
(d\lambda_{nm}^{\varepsilon\delta}+k-i\delta m\mu)}{16\pi^2nm}.
\ee

According to the experimental data \cite{Ak02,KOP03,MGB06,GAM06,GS06}
the frequency of sound emitted by a singing wine glass and a squealing laboratory brake at low spins
is close to a double
eigenfrequency of the non-rotating bodies.
For this reason it is important to study first the influence of external stiffness,
damping and friction on the deformation
of the spectral mesh near the nodes corresponding to $\Omega=0$.

    \begin{figure}
    \begin{center}
    \includegraphics[angle=0, width=0.65
    \textwidth]{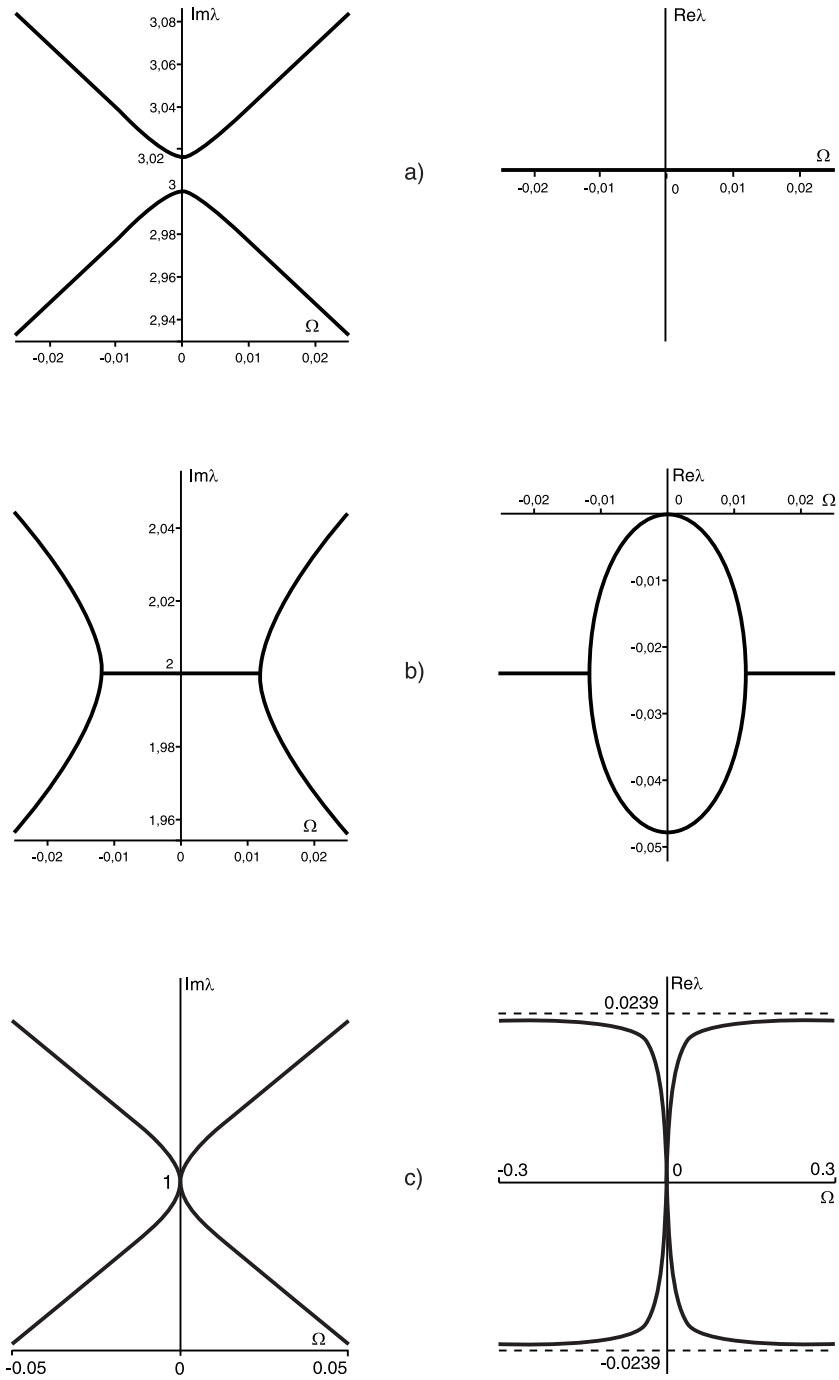}
    \end{center}
    \caption{Deformation of the spectral mesh of the rotating string near the nodes $(0,3)$, $(0,2)$, and $(0,1)$
    caused by the action of the external spring with $k=0.3$ (a), damper with $d=0.3$ (b), and friction with $\mu=0.3$ (c),
    respectively.}
    \label{fig7}
    \end{figure}

At $\Omega=0$ the double eigenvalue $\lambda_{nm}^{\varepsilon\delta}$ originates due to
intersection of the eigenvalue branches $\lambda_n^{\varepsilon}$ and $\lambda_m^{\delta}$,
where $m=n$ and $\varepsilon=-\delta$. Taking this into account, we find from \rf{s19} and \rf{s20}
the expressions describing splitting of the double eigenvalue $in$ due to action of gyroscopic forces
and an external spring
\be{s21}
\lambda=in+i\frac{k}{4\pi n}\pm i\sqrt{n^2\Omega^2+\frac{k^2}{16\pi^2n^2}}.
\ee
Equation \rf{s21} describes a hyperbola in the plane ${\rm Im}\lambda$ versus $\Omega$,
which intersects the axis $\Omega=0$ at the values $\lambda=n$ and $\lambda=n+\frac{k}{2\pi n}$, see Fig.~\ref{fig7}(a).
The gap between the branches decreases with the increase of the number $n$ of a mode.
It is remarkable that the lower branch passes through the point corresponding to the node of the
spectral mesh of the non-perturbed gyroscopic system, which agrees with the numerical results of \cite{YH95}.
Remember that in case of two-dimensional systems
the reason for such a degenerate behavior is a zero eigenvalue in the matrix $\bf K$ of external potential fores.

Analogously, from expressions \rf{s19} and \rf{s20} follows the asymptotic formula for the eigenvalues
originating after the splitting of the double eigenvalue $in$ at $\Omega=0$ due to perturbation by the
gyroscopic forces and an external damper
\be{s22}
\lambda=in-\frac{d}{4\pi}\pm\sqrt{\frac{d^2}{16\pi^2}-n^2\Omega^2}.
\ee
According to \rf{s22} the real parts of the eigenvalues as functions of $\Omega$ originate a
bubble of instability in the plane ${\rm Re}\lambda$ versus $\Omega$
\be{s23}
\left({\rm Re}\lambda+\frac{d}{4\pi} \right)^2+n^2\Omega^2=\frac{d^2}{16\pi^2},\quad {\rm Im}\lambda=n.
\ee
The ellipse \rf{s23} is submerged under the plane ${\rm Re}\lambda=0$ in the space
$(\Omega, {\rm Im}\lambda, {\rm Re}\lambda)$ so that it touches the plane at the origin, as shown
in Fig.~\ref{fig7}(b). The ellipse is connected with the branches of the hyperbola of complex eigenvalues
\be{s24}
n^2\Omega^2-\left( {\rm Im}\lambda - n \right)^2=\frac{d^2}{16\pi^2},\quad
{\rm Re}\lambda=-\frac{d}{4\pi}.
\ee
The horizontal diameter of the ellipse decreases with the increase in the mode number $n$, while
the vertical one does not change.
As we see, the external damper creates a latent source of local subcritical flutter instability exactly as it happens
in two dimensions when the matrix of dissipative forces $\bf D$ is semi-definite, i.e. when it has one zero eigenvalue.
The range of change of the gyroscopic parameter corresponding to the latent instability
is located compactly around the origin and decreases with the increase in $n$.

The deformation of the spectral mesh near the double eigenvalue $in$ at $\Omega=0$ due to combined action of
gyroscopic forces and external friction is described by the expression
\be{s25}
\lambda=in\pm\sqrt{\left(in\Omega+\frac{\mu}{4\pi}\right)^2-\frac{\mu^2}{16\pi^2}},
\ee
following from \rf{s19} and \rf{s20}.

The imaginary parts of the eigenvalues of the deformed spectral mesh
\be{s26}
{\rm Im}\lambda=n\pm\frac{1}{2\pi}\sqrt{2\pi^2 n^2\Omega^2\pm\pi n\Omega \sqrt{4\pi^2n^2\Omega^2+\mu^2}}
\ee
cross at the node $(0,n)$, as in the non-perturbed case. However, the crossing is degenerate because
the eigenvalue branches touch each other at the node, see Fig.~\ref{fig7}(c). Indeed, expanding expression \rf{s26}
in the vicinity
of $\Omega=0$ we find that
\be{s27}
{\rm Im}\lambda=n\pm\frac{1}{2\pi}\sqrt{\pi n \mu |\Omega|}+O(\Omega^{3/2}).
\ee
Clearly, at $\Omega=0$ the imaginary parts do not split due to non-conservative perturbation from the eyelet.
It is instructive to note that for $\Omega \rightarrow \infty$ the imaginary parts asymptotically tend to
$n(1\pm \Omega)$. By this reason for small perturbations the spectral mesh depicted in the plane ${\rm Im}\lambda$
versus $\Omega$ looks non-deformed at the first glance.

The real parts of the deformed spectral mesh described by the expression
\be{s28}
{\rm Re}\lambda=\pm\frac{1}{2\pi}\sqrt{-2\pi^2 n^2\Omega^2\pm\pi n\Omega\sqrt{4\pi^2n^2\Omega^2+\mu^2}}
\ee
cross at $\Omega=0$, so that ${\rm Re}\lambda=0$.
The crossing of the real parts is degenerate Fig.~\ref{fig7}(c), which is confirmed by the asymptotic expression
\be{s29}
{\rm Re}\lambda=\pm\frac{1}{2\pi}\sqrt{\pi n \mu |\Omega|}+O(\Omega^{3/2}).
\ee
For $\Omega \rightarrow \infty$ the real parts follow the asymptotic law
\be{s30}
{\rm Re}\lambda=\pm\frac{\mu}{4\pi}\mp\frac{\mu^3}{128\pi^3n^2\Omega^2}+o(\Omega^{-2})
\ee
As is seen in Fig.~\ref{fig7}(c), the real parts almost always are close to the lines $\pm \mu/(4\pi)$,
except for the vicinity of the node of the spectral mesh, where the real parts rapidly tend to zero.
This behavior agrees with the results of numerical calculations of \cite{YH95}.

We see that the double semi-simple eigenvalue $in$ does not split
due to variation of only the parameter of non-conservative positional forces $\mu$.
In the two-dimensional case this would correspond to the degenerate matrix $\bf N$, $\det{\bf N}=0$.
Since $\bf N$ is skew-symmetric, the degeneracy means ${\bf N}\equiv 0$, i.e. the absence of non-conservative perturbation.
In case of more then two degrees of freedom the degeneracy of the operator of non-conservative positional forces
leads to the cuspidal deviation of the generic splitting picture registered in Fig.~\ref{fig7}(c).

We have considered local deformation of the spectral mesh of the rotating string
near the nodes at $\Omega=0$. However, the formulae \rf{s19} and \rf{s20} enable us to
investigate similarly the deformation of the mesh in the vicinity of all the other nodes.

Concluding, we note that the source of perturbation of the rotating string concentrated at one point
leads to the deformations of its spectral mesh, which correspond to that caused by the semi-definite matrices of
conservative, dissipative and non-conservative forces in the two-dimensional case.
Zero eigenvalues of the damping
operator encourage the activation of the latent bubble of instability, while the zero eigenvalue of the operator
of non-conservative positional forces suppresses this process. To get local subcritical flutter instability,
described in the previous sections for the finite-dimensional model, the operators
of dissipative and non-conservative perturbations must be generic, which excludes their semi-definiteness.
We notice that the singular sources of momentum
and energy in the problem of stability of a rotating string as a reason for degeneracies were discussed also in \cite{OV03, OV04}.
One of the ways to avoid this degeneracy is to consider not pointwise but distributed contact with the dissipative,
stiffness and friction characteristics depending on the material coordinates. A step in this direction is
taken in \cite{SHKH07}, where a model of distributed pads was developed. For simplicity, in \cite{SHKH07}
the characteristics of the
pins constituting the pads were assumed not depending on the coordinates,
which led to the same semi-definite degeneracy.
This could be a reflection of the so-called Herrmann-Smith paradox of a beam resting on a Winkler-type elastic foundation
and loaded by a follower force \cite{KS02}. The degeneracy in the Herrmann-Smith problem is removed by assuming a
non-uniform modulus of elasticity. Similar modification of the model of the distributed brake pads of \cite{SHKH07}
could give generic perturbation operators and open the way to the modeling of the disk brake squeal catching
its most significant features.

\section{Conclusion: Disc brake as a musical instrument}

As we already mentioned, the principle of activating sound by friction is the same for a wine glass, a disc brake, and
the glass harmonica. The latter is an ancient musical instrument for which
the famous "Dance of the Sugar Plum Fairy" in the first edition of "The Nutcracker" ballet was composed
by P.I. Tchaikowsky in 1891 \cite{GA}.

The results obtained in the present paper show that the "keyboard" of rotating elastic bodies of revolution,
among which are the glass harmonica and the disc of a brake, is formed by the nodes of the spectral mesh, corresponding
to angular velocities in the subcritical range.
The frictional contact is the source of dissipative and non-conservative forces, which
make the system unstable in the vicinity of the nodes and force a rotating structure to vibrate at
a frequency close to the double frequency of the node and at the angular velocity close to that
of the node. These conclusions agree with the results of recent experiments with the laboratory brake
\cite{MGB06,GAM06,GS06}.
The higher modes are more efficiently damped than the lower ones, and
a particular frequency is selected by the speed of rotation and the loading conditions, including such parameters
as the size of the friction pads and their placement with respect to the disc.

It is known that dissipative and non-conservative forces may influence the stability in a non-intuitive manner
\cite{La75,Ka75,MK91,BR94,HG03,M04,Ma06,Ki06,Ki07,KM07}.
We have shown that the former create the latent local sources of instability around the nodes of the spectral mesh
(bubbles of instability),
while the latter activate these sources by inflating and destructing the bubbles.
It turns out that the eigenvalues of the damping matrix control the development of instability.
For better stability both of them should be positive and stay far from zero. If one of the eigenvalues of
the damping matrix is close to zero or becomes negative, the instability can occur with the weaker non-conservative
positional forces or even without them.

With the use of the perturbation theory of multiple eigenvalues we have obtained explicit formulas describing the
deformation of the spectral mesh by dissipative and non-conservative perturbations. The trajectories of eigenvalues
are analytically described and classified. The approximations of the domain of asymptotic stability are obtained
with the use of the derivatives of the operator and the eigenvectors of the double eigenvalues calculated at the
nodes of the spectral mesh. Singularities of the stability boundary of a new type were found and their role
in the development of instability was clarified.

The theory developed seems to be the first analytical explanation of the basic mechanism of
friction-induced instabilities in rotating elastic bodies of revolution. The application of the theory
to more complicated discrete and continuous models
of squealing disc brakes, singing wine glasses, and calender rolls will be published elsewhere.
Comparing the generic results obtained in the two dimensional case with that of the study of the rotating string
with the singular source of momentum and energy we have shown that the traditionally employed concept
of a point-wise contact leads to the semi-definite non-generic perturbation operators which suppress generic instability
mechanism causing squeal. The further progress in the squeal simulation seems to strongly depend on creation of the models
of distributed contact that do not allow for the semi-definite degeneration.

\section*{Acknowledgements}
The work has been supported by the Alexander von Humboldt
Foundation.
\appendix

\end{document}